\documentclass[journal,twoside,web]{IEEEtran}
\usepackage{generic}
\usepackage{cite}
\usepackage{amsmath,amssymb,amsfonts}
\usepackage{algorithmic}
\usepackage{graphicx}
\usepackage{algorithm,algorithmic}
\usepackage{hyperref}
\usepackage{autobreak}
\hypersetup{hidelinks=true}
\usepackage{textcomp}
\usepackage[utf8]{inputenc}
\usepackage{graphicx}
\graphicspath{{Images/}}
\def\BibTeX{{\rm B\kern-.05em{\sc i\kern-.025em b}\kern-.08em
    T\kern-.1667em\lower.7ex\hbox{E}\kern-.125emX}}
\begin{document}
\title{Thoracic Fluid Measurements by Bioimpedance: A Comprehensive Survey}
\author{Manender Yadav$^{1,2}$, Shreyansh Shukla$^3$, Varsha Kiron$^4$, U. Deva Priyakumar$^2$, Maitreya Maity$^1$
\thanks{This work is funded by $^1$IHub-data, International Institute of Information Technology Hyderabad}
\thanks{$^2$Centre For Computational Natural Sciences and Bioinformatics, International Institute of Information Technology, Hyderabad- 500032 }
\thanks{$^3$Univerity of Mumbai-Department of Atomic Energy-Centre for excellence in basic Sciences, Mumbai}
\thanks{$^4$AIG Hospital Banjara hills (8-2-594/FF, Zahara Nagar, Banjara hills) Hyderabad}
}

\maketitle

\begin{abstract}
Bioimpedance is an extensively studied non-invasive technique with diverse applications in biomedicine. This comprehensive review delves into the foundational concepts, technical intricacies, and practical implementations of bioimpedance. It elucidates the underlying principles governing bioimpedance measurements, including the relevant physics equations employed for estimating body fluid levels. Moreover, a thorough examination of the prevalent single-chip analog front end (AFE) available in the market, such as AD5933, MAX30001, AD5940, and AFE4300, is conducted, shedding light on their specifications and functionalities. The review focuses on using bioimpedance to assess thoracic impedance for heart failure detection by utilizing the relation between lung water and heart failure. Traditional techniques are compared with bioimpedance-based methods, demonstrating the latter's efficacy as a non-invasive tool for cardiac evaluation. In addition, the review addresses the technical limitations and challenges associated with bioimpedance. Pertinent issues such as contact impedance, motion artifacts, calibration, and validation regarding their impact on measurement precision and dependability are thoroughly examined. The review also explores strategies and advancements in using artificial intelligence to mitigate these challenges.
\end{abstract}

\begin{IEEEkeywords}
Bioimpedance, Thoracic fluid, Total body water, Body composition, Heart failure, Biosensors, Analog front end.   
\end{IEEEkeywords}

\section{Introduction}
\label{sec:introduction}
The field of bioimpedance has been making huge progress recently, revolutionizing the way we assess various physiological parameters \cite{thanapholsart2023current}. Bioimpedance-based methods, which are non-invasive and quick, have been receiving increased interest in both clinical practice and research studies. Electrical properties of tissues and bioimpedance provide valuable insights into the physiological characteristics of the body, e.g., water, fat, protein, etc. \cite{piccoli1994new}. Cardiovascular diseases, which caused 20.5 million fatalities in 2021, are the leading cause of mortality in the whole world. The most common cause of thoracic fluid accumulation is congestive heart failure (CHF). Traditionally, the detection of thoracic fluid relied on invasive and time-consuming techniques such as pathological tests, CT scans, MRI, X-rays, etc. \cite{wabel2008towards}. However, bioimpedance can play a major role in measuring and analyzing thoracic fluid content more efficiently. This comprehensive systematic literature review article explores the fundamentals, theory, and concepts of bioimpedance, with a particular focus on the advancements made in single-chip sensors and their application in thoracic bioimpedance measurement. 
\par This systematic literature review (SLR) begins with a detailed description of the article selection method section \ref{sec:asm}, which outlines the research questions guiding this article, the methodological procedures used for paper selection, and a comprehensive list of the databases searched for relevant research papers. The next section \ref{sec:bb}, Bioimpedance basics, is devoted to explaining the foundational principles of bioimpedance. This section provides a thorough exploration of the theoretical underpinnings, describing the interaction of electrical currents with biological tissues and its role in generating valuable information. The following section \ref{sec:cd}, Chips and devices, focuses on technological advancements in single-chips. These innovations have significantly improved the practicality and versatility of bioimpedance measurements. Technical specifications of prominent single-chip sensors are presented, along with an extensive examination of studies employing these devices. The main emphasis of this article is on thoracic fluid detection, which is a practical application of bioimpedance technology. This topic is comprehensively addressed in the dedicated application section \ref{sec:app}. Here, the authors objectively assess how bio-impedance-based techniques compare with conventional methods, elucidating their inherent strengths, limitations, and potential for further improvement. Their analysis is based on a rigorous evaluation of accuracy, reliability, and efficiency. Next, the authors offer an extensive exploration of the integration of Artificial Intelligence (AI) within the realm of bioimpedance in section \ref{sec:ai}. This section illustrates how AI enhances the capabilities of bioimpedance technology in various applications, with an emphasis on the underlying scientific principles. In the subsequent section \ref{sec:chal}, the authors address the major challenges faced by the researchers in this field, acknowledging the inherent limitations of bioimpedance technology. Finally, we concluded the review work in section \ref{sec:con}.
\par Overall, this review article aims to serve as a valuable resource for researchers, clinicians, and healthcare professionals interested in the field of bioimpedance. By presenting a detailed analysis of bioimpedance basics, theory, and concepts, as well as the latest single-chip sensors available in the market, we hope to foster a deeper understanding of this technology and its applications in thoracic fluid detection.
\section{Article selection method}
\label{sec:asm}
\begin{figure*}
    \centering
    \includegraphics[width=\textwidth]{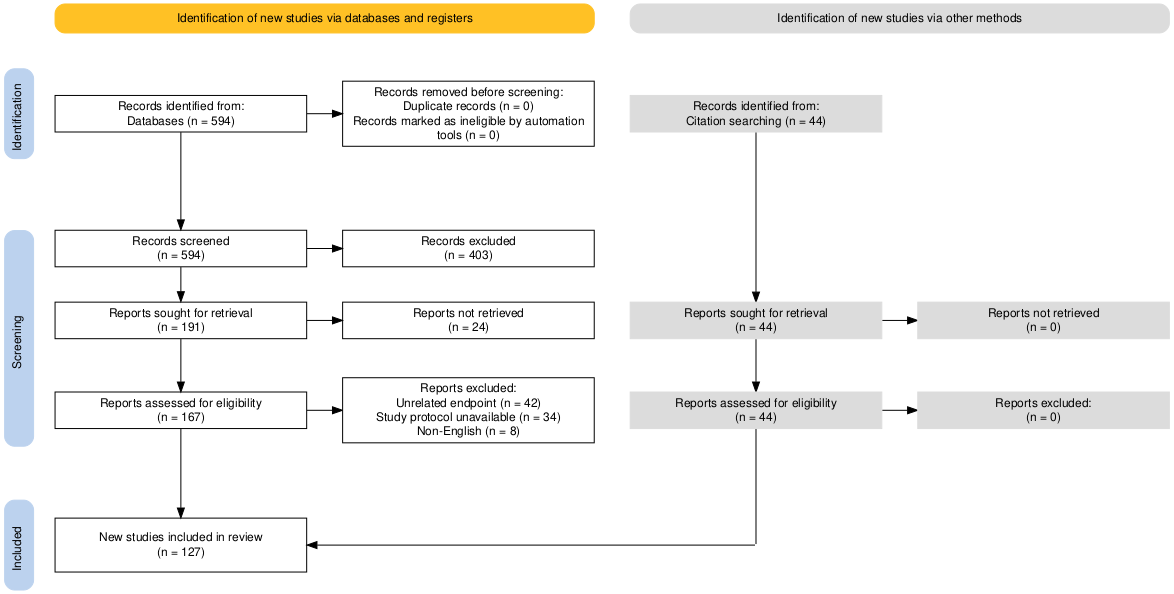}
    \caption{Article selection based on PRISMA framework.}
    \label{fig: Prisma}
\end{figure*}

A systematic literature review (SLR) is a research methodology that employs a structured and comprehensive approach to identify, select, and evaluate relevant studies to answer a specific research question\cite{okoli2015guide}. The methodology involves formulating research questions, defining inclusion and exclusion criteria, and conducting a thorough search of multiple databases. In this review, the authors conducted a comprehensive search of several databases supported by Google Scholar, including IEEE, Science Direct, Springer, Hindawi, and PubMed. The search spanned from 2015 to 2025 and included all relevant publications published within this period. The authors used the following keywords to search for relevant publications: bioimpedance, body composition, fluid management, health monitoring, heart failure, impedance analyzers, machine learning, artificial intelligence, single-chip biosensors, thoracic fluid, tissue impedance, and total body water. The search strategy yielded 594 articles that were screened for relevance based on their abstracts as shown in Fig.\ref{fig: Prisma}. The selected articles were then thoroughly studied to identify important papers which were cited but were outside the search year and were included in this review. Finally, 127 articles were included in this comprehensive review article.

\begin{figure}
    \centering
    \includegraphics[width=8cm]{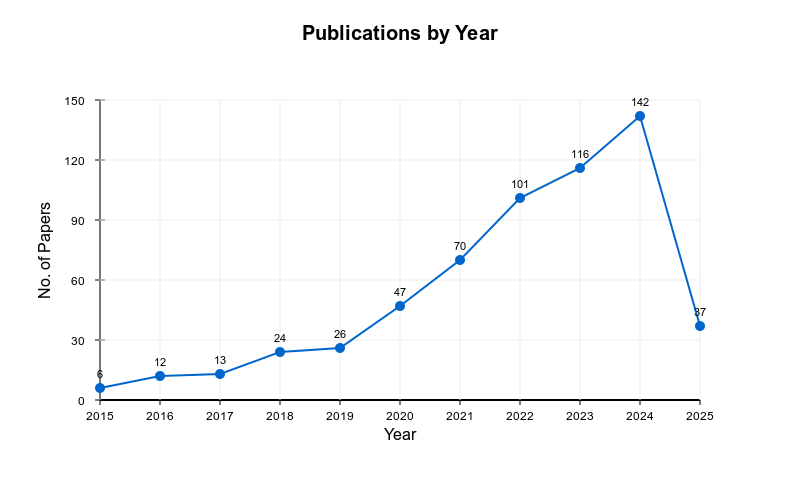}
    \caption{Articles based on the above-mentioned keywords by year}
    \label{fig: paper by year}
\end{figure}
The number of articles published in this domain for the past 10 years is shown in Fig.\ref{fig: paper by year}. There is a gradual increase in the number of papers published every year related to this domain, the sudden decrease in 2025 is likely because the data only represents approximately the first quarter of the year. When analyzed in depth, the recent papers were found to be more relevant and involved the use of Machine Learning methodologies. This signifies a shift from the traditional methods to the more advanced AI-based research solutions. The following sections discuss the findings of this in-depth literature review.
\section{Bioimpedance basics}
\label{sec:bb}
Bioimpedance refers to how biological systems or tissues interact with electricity. A small potential difference is applied to the biological tissue and measurement of the resulting electrical signals is analyzed to obtain various physiological parameters. Bioimpedance measurements are typically performed using electrodes placed on the skin's surface or by utilizing specialized sensors. Multi-frequency current is applied to the tissue of interest, depending on the specific application. The impedance of biological tissues arises from various components, including cells, extracellular fluid, blood vessels, and cell membranes. These components have different electrical properties, such as resistance (R) and reactance (X), which contribute to the overall impedance. Bioimpedance analysis (BIA) is commonly used to estimate body composition parameters. In BIA, the impedance of the body is measured typically by passing a low-intensity AC current and then derives information about body fat, lean mass, hydration status, and other relevant parameters. Another application of bioimpedance is electrical impedance tomography (EIT), which involves measuring impedance changes at multiple points on the body's surface. EIT can provide real-time imaging of the electrical properties within the body, making it useful for monitoring lung function, analyzing brain activity, and other medical applications. Bioimpedance has found applications in various fields, including medicine, sports science, and nutrition. It offers non-invasive or minimally invasive ways to gather information about physiological conditions and can aid in diagnosing certain diseases, assessing body composition, monitoring hydration levels, and guiding therapeutic interventions.
\par Bioelectrical impedance analysis (BIA)\cite{mialich2014analysis} and bioelectrical impedance spectroscopy (BIS)\cite{barbosa2005bioelectrical} are techniques used to study tissues, providing valuable insights into their molecular composition and physical structure for diagnostic and prognostic purposes. The human body along with the organs and cells, exhibit structural heterogeneity, and their molecular composition varies due to dynamic biochemical reactions and the presence of electro-responsive components like ions, proteins, and polarized membranes. Despite the challenges in interpreting bioimpedance data, BIA/BIS has proven to be a powerful tool for evaluation and monitoring \cite{abasi2022bioelectrical}. BIA/BIS serves as the foundation for various diagnostic and monitoring systems. Its popularity can be attributed to three key factors:
\begin{enumerate}
  \item It is a non-invasive and safe measurement method
  \item it can be easily miniaturized
  \item many technological formats exist for its implementation in biomedicine \cite{annus2021bioimpedance}
\end{enumerate}
\subsection{Principal behind bioimpedance}
Tissues are composed of cells and membranes, which are thin and have a high resistivity, functioning like capacitors. When high frequencies are used, the current can pass through these capacitors, and the outcome is influenced by both the tissue and the fluid inside and outside it. At lower frequencies, the membranes obstruct current flow, providing information only about the external fluid. Fat has a high resistivity, while blood reduces the resistivity.
\par Bioimpedance techniques employ electrodes that have a galvanic connection to the tissue. Bioimpedance measurement devices use electronic circuits and electrodes. In copper wires, electrons carry a charge, while in tissue, ions (mostly) carry the charge. The electrode is where the conversion of charge carriers from ions to electrons and vice versa occurs.\\
Now let's look at the equations for bioimpedance,
\begin{equation}
Z=\rho L/A
\end{equation}
where $\rho$ is resistivity of the tissue, Z denotes the impedance, L and A being the electrodes' distance and cross-section area, respectively.\\ 
The above model is only for the homogeneous tissue and current band electrodes are far away from the pickup electrodes\\\cite{thompson2019quantification}.

\subsection{Bio-impedance equations from Maxwell's equations}
 Relevant Maxwell equation most for bioimpedance is \cite{bari2018electrodermal}:
\begin{equation}
\begin{aligned}
 \nabla \times \mathbf{H} & = \frac{\partial \mathbf{D}}{\partial t}
  & = \mathbf{J}   
\end{aligned}
\end{equation}
where \(\mathbf{H}\) is magnetic field [A/m], \(\mathbf{D}\) denotes the electric flux density [coulomb/m\(^2\)].\\
The electric flux density \(\mathbf{D}\) can be written as:
\begin{equation}
\mathbf{D} = \varepsilon_0 \mathbf{E} + \mathbf{P} 
\end{equation}
where \(\mathbf{E}\) denotes the electric field [V/m], \(\varepsilon_0\) is the permittivity of vacuum [farad (F)/m] and electric polarization is represented by \(\mathbf{P}\)(dipole moment per volume [coulomb/m\(^2\)]).\\
When the magnetic component is ignored, Equation (2) becomes:
\begin{equation}
    \frac{\partial \mathbf{D}}{\partial t} =-\mathbf{J}= \sigma \mathbf{E} 
\end{equation}
where \(\sigma\) is the conductivity [S/m].\\
We assume a small voltage amplitude (V) across the material, thus making so the system linear. As sinusoidal functions in input, we can represent it with complex notation as a derivative. Use of $D = eE$ (space vectors), where the permittivity $e = e_r e_o$ under these conditions, lossy dielectric can be characterized by a complex dielectric constant,
\begin{equation}
\epsilon=\epsilon'-\mathbf{i} \epsilon"\end{equation}
same for conductivity 
\begin{equation}
\sigma= \sigma'+\mathbf{i} \sigma"\end{equation}
where, $\mathbf{i}=\sqrt{-1}$ .
using this and applying equation 2 on a capacitor of area A and dielectric thickness L. But equation (2) is in a differential form, and the discontinuous interface will have discontinuity thus, we will use Gauss Law as an integral form.
The equations above are not just extremely robust but are valid under non-homogeneous, nonlinear, and anisotropic conditions. The bio-impedance equation, which is derived from the Maxwell equation, is given by:
\begin{equation}
Y = G + \mathbf{i}\omega C 
\end{equation} 
In which (Y) denotes the admittance [siemens, S], (G) symbolizes the conductance [S], ( $\omega$ ) represents the angular frequency [rad/s], (C) is the capacitance [farad, F] and $\mathbf{i}=\sqrt{-1}$. The equation for impedance reveals that the fundamental impedance model is essentially a model of admittance. In this model, the conductive and capacitive components are physically parallel. It’s crucial to note that this model primarily depicts a dielectric model with dry samples. \\
We have obtained an impedance equation for a circuit. This makes the next question about what is impedance. It is the effective resistance of an electrical circuit to alternating current (ohmic resistance and reactance); to be more specific, it is the resistance a circuit or components offers to current flow in a circuit at a specific frequency unit ohms ($\Omega$). It changes due to the reactive components non-reactive - resistor (same for AC and DC current) reactive components - mainly capacitor and inductor.
\subsubsection{Capacitor}The impedance of this reactive component displays an inverse relationship with frequency, being substantial at low frequencies and minimal at high frequencies. Capacitors store an electrical charge. The capacitor charges and discharges to resist change in the current. In a capacitor, the faster the direction changes - the lower the capacitive resistance.
\subsubsection{Inductor}The impedance of this reactive device displays a direct relationship with frequency, being minimal at low frequencies and significant at high frequencies. Coils and reactors are common sources of inductive resistance. These components establish a magnetic field that opposes changes in the direction of alternating current. Specifically, in an inductor, the faster the AC current's direction shifts, the higher the inductive resistance it exhibits.
\subsubsection{Formula governing capacitive impedance}
The formula for impedance in ohm $\Omega$\cite{showkat2023review}
\begin{equation}{X_c = \frac{1}{2*\pi f * C}}\end{equation}
Formula governing inductive impedance\\
The formula for inductive impedance in ohm $\Omega$
$$X_L = 2\pi *f * L $$
total reactive impedance is given by 
$$ X_{total} = X_c + X_L $$
\begin{equation}
\begin{aligned}
    X_{total} = \frac{1}{2\pi* f * X_c} + 2\pi *f * X_L
\end{aligned}
\end{equation}
total impedance is given by 
\begin{equation}Z=R+\mathbf{i}X_{total}\end{equation} 
where, $f$ is frequency, $X_c$ is Capacitance, $R$ is non reactive resistance and $X_L$ Inductance thus making the magnitude detected by the meter as

\begin{equation}
\begin{split}
Z & =\sqrt{R^2+X_{total}^2} \\
& =\sqrt{R^2+|X_L-X_c|^2}
\end{split}
\end{equation}
Where R is the nonreactive impedance or simple resistance.
\begin{equation}\omega=2\pi * f\end{equation}
Thus, for ease, it can be replaced in the above equations. To measure impedance in the physics lab, we used a lock-in amplifier.\\
To measure different impedance\\
Ohm's law: relation with potential difference and current
\begin{equation}R=\frac{\Delta V}{I}\end{equation}
unit 
\begin{equation}\Omega = \frac{V}{A}\end{equation}
for  2 terminal circuits, it is the ratio of the complex representation of the sinusoidal voltage difference. For sinusoidal waves, we have the following equations:
\begin{equation}V=|V|e^{\mathbf{i}(wt+\phi_V)}\end{equation}
\begin{equation}I=|I|e^{\mathbf{i}(wt+\phi_I)}\end{equation}
and,
\begin{equation}Z= \frac{V}{I}\end{equation}
One point to note is that the inductor leads by a phase of $\pi/2$, and the capacitor lags by the same.
\subsection{Dependence on frequency and input signal }
The way biological tissue conducts electricity is affected by the frequency of the signal it is exposed to. Tissue can be modeled as either a conductor or a dielectric, depending on the frequency range. Below 100 kHz, tissue behaves mostly as an electrolytic conductor, with capacitive effects due to electrode-electrolyte interfaces and cell membranes. These effects can be measured with high-resolution techniques, such as lock-in amplifiers, to reveal the dielectric properties of tissue at low frequencies. Above 50 kHz, the tissue becomes more dielectric, with its properties approaching those of water at very high frequencies (around 18 GHz). Water has a characteristic relaxation frequency where its dielectric loss factor is maximal.

To measure the bioimpedance or bioimmittance of tissue, different types of waveforms and electrode configurations can be used. Sinusoidal waveforms are preferred to overstep or pulse waveforms, as they avoid nonlinearities and heat generation that may alter the tissue properties. Single-frequency (BIA) or multi-frequency (BIS) signals can be injected into the tissue with low amplitude and measured with a voltage meter.

The impedance varies with frequency, showing two main dispersions: alpha and beta. Alpha dispersion occurs when the applied signal frequency is in between 10 Hz to 10 kHz and is related to the membrane polarization of cells. Beta dispersion occurs when it is in between 10 kHz to 10 MHz and is related to the interfacial polarization of tissue components. Two other dispersions, delta and gamma, are less relevant for bioimpedance measurements, as they occur at higher frequencies (above 100 MHz) and are influenced by water molecules and macromolecules.

To measure bioimpedance accurately, a four-electrode (tetrapolar) system is recommended, as it reduces the contact resistance errors that may arise from using two electrodes for both current injection and voltage sensing. A four-electrode system consists of two pair of electrodes, one pair is used to apply the current and the other pair to measure the voltage. The transfer impedance is calculated by dividing the measured voltage by the applied current. If no voltage is measured, the transfer impedance is zero, which means that no signal transfer occurs between the electrodes.

Bioimpedance measurements can be used to analyze and infer biological events and underlying principles, such as tissue composition, hydration status, blood flow, cell viability, etc. Bioimpedance can also be integrated into wearable devices and connected to the Internet for remote monitoring and diagnosis. To extract relevant physiological or compositional information from bioimpedance data, equivalent circuit models are used to represent the electrical properties of tissue. Some examples of circuit models are the Randles circuit, the Cole-Cole circuit, and the Cole equation. These models use different combinations of resistors, capacitors, constant phase elements, and Warburg impedances to fit the impedance data and estimate parameters such as resistance, reactance, phase angle, etc\cite{martinsen2014bioimpedance}
\cite{abasi2022bioelectrical}\cite{abraham2007intrathoracic}\cite{jaffrin2008body}.

\begin{table*}
\centering
\caption{Total body water prediction equations}
\label{tab:tbw equations}
\resizebox{\textwidth}{!}{%
\begin{tabular}{|c|c|c|c|c|}
\hline
\textbf{Author} & \textbf{Equation} & \textbf{Frequency} & \textbf{Reference}\\
\hline
Divala et.al. & $\text{TBW (kg)} = 0.328 \times \frac{\text{Height}^2 \text{(cm)}}{\text{Resistance (}\Omega\text{)}} + 0.910 \times \text{sex}+ 0.307 \times \text{age (years)} + 0.249 \times \text{Weight} + 0.015 \times \text{Reactance (}\Omega\text{)}-2.152$& 50 KHz & \cite{divala2022development}\\
\hline
Hannan et.al. & $\text{TBW (kg)} = 0.2391 \times \frac{\text{Height}^2 \text{(cm)}}{\text{Resistance (}\Omega\text{)}} + 2.971 \times \text{sex} + 0.1889 \times Weight + 5.4641$ & 200 KHz & \cite{hannan1995comparison}\\
\hline
Sun et.al. & $\text{TBW (kg) male} = 0.45 \times \frac{\text{Height}^2 \text{(cm)}}{\text{Resistance (}\Omega\text{)}} + 0.18 \times Weight + 1.20$ & 50 KHz & \cite{sun2003development}\\
\hline
Sun et.al. & $\text{TBW (kg) female} = 0.45 \times \frac{\text{Height}^2 \text{(cm)}}{\text{Resistance (}\Omega\text{)}} + 0.11 \times Weight + 3.75$ & 50 KHz & \cite{sun2003development}\\
\hline
Kushner et.al. & $\text{TBW (kg)} = 0.5561 \times \frac{\text{Height}^2 \text{(cm)}}{\text{Resistance (}\Omega\text{)}} + 0.0955 \times Weight + 1.726$ & 50 KHz & \cite{kushner1986estimation}\\
\hline
\end{tabular}%
}
\end{table*}

\subsubsection{Contributions to Ionic Conductivity}
Kohlrausch demonstrated that the conductivity, represented as $\sigma$, is the aggregation of individual contributions from anions (-) and cations (+). The current density, denoted as J [A/m$^2$], for a single pair of anion-cation, is given by:
\begin{equation}J=(nzev)+ +(nzev)- = \sigma E [A/m^2]\end{equation}
\begin{equation}\sigma=Fc\gamma(\mu_- +\mu_+) [S/m]\end{equation}
In these equations, n is the ion concentration per volume, z is the valency of an atom (the number of transferable electrons), e is the charge of the electron [C], and v is the ion velocity [m/s]. F represents the Faraday constant [C/mol], c is the concentration [mol/m3], g is the activity coefficient (which accounts for possible incomplete electrolyte dissociation, with g ranging between 0 and 1), and m signifies the mobility [m2/Vs].
The current density, J, must include contributions from each species of negative and positive ions. It’s important to note that determining individual ion species’ activity coefficients can be challenging due to electroneutrality - an electrolyte cannot consist solely of anions or cations. From this, we can then compute total conductivity and molar conductivity. It’s also worth noting that friction (as per Stokes’ law f=6$\pi$$\nu$av) will also be a factor.
\subsection{Sensitivity and geometry of electrode}
The electrical behavior of biological tissues in bioimpedance analysis is often represented through electrical circuit models, a key aspect of the network analysis and synthesis method.The following models can be used to analyze and interpret bioimpedance measurements.
\subsubsection{Randles circuit}
One of the most commonly used circuit models in bioimpedance is the Randles circuit. The model incorporates a resistor denoted by (R), a constant phase element (CPE) as its constituent parts. The Randles circuit is given by:
\begin{equation}
Z_{Rand} = R + \frac{1}{Q(CPE(\mathbf{i}\omega)^{\alpha})} + W(\mathbf{i}\omega)^{-1/2} 
\end{equation}
Where \(Z_{Rand}\) is the impedance of the Randles circuit, \(Q\) is the constant phase element, \(\alpha\) is the Warburg coefficient, and \(W\) is the Warburg impedance\cite{laschuk2021reducing}.
\subsubsection{Cole-Cole circuit}
Another commonly used circuit model is the Cole-Cole circuit, which is given by:
\begin{equation}
Z_{CC} = R + \frac{1}{Q(CPE(\mathbf{i}\omega)^{\alpha})} 
\end{equation}
Where \(Z_{CC}\) is the impedance of the Cole-Cole circuit\cite{stroud1995use}\cite{lazovic2014modeling}.\\
The Cole equation is a simplified version of the Cole-Cole circuit, which assumes that the constant phase element is a pure capacitance. The Cole equation is given by:
\begin{equation}
Z_{Cole} = R + \frac{1}{Q(\mathbf{i}\omega)^{\alpha}} 
\end{equation}
where \(Z_{Cole}\) is the impedance of the Cole equation.\\
In addition to these circuit models, bio-impedance measurements can also be analyzed to model the complex dielectric response of biological tissues using equivalent circuit models based on the how the relaxation times are distributed.\\
Overall, the network analysis and synthesis approach provides a powerful tool for analyzing and interpreting bio-impedance measurements. By using these electrical circuit models, researchers can understand the underlying physiological processes that give rise to bio-impedance measurements.\\
One important property is the resistivity $R_{total}$ of the tissue, which represents its inherent resistance to the flow of electric current. The resistivity can be complex, denoted as 
\begin{equation}R_{total} = R + \mathbf{i}(R_L+R_C)\end{equation}
where \(R\) denotes the real part and \(R_L+R_C\) denotes the imaginary part.\\
The conductivity $\sigma$ of the tissue is the reciprocal of the resistivity and is given by
\begin{equation}\sigma = 1/R_{total}\end{equation}
It represents the ease with which electric current flows through the tissue.\\
The permittivity $\varepsilon$ of the tissue describes its ability to store electrical energy in an electric field. It can also be complex, denoted as 
$$\varepsilon = \varepsilon' + \mathbf{i}\varepsilon''$$
where $\varepsilon'$ and $\varepsilon''$ are the real and imaginary part respectively.\\
The complex permittivity can be related to the complex conductivity through the equation 
\begin{equation}\varepsilon = \varepsilon_0 \sigma\end{equation} 
where \(\varepsilon_0\) denotes the permittivity of vacuum as a constant.\\
Similarly, the impedance $z$ of the tissue is also the opposition to the flow of electric current and is given by 
\begin{equation}
    Z = R_{total} + \mathbf{i} \omega \varepsilon
\end{equation}
where $\omega$is the angular frequency.\\
The admittance $Y$ of the tissue is the reciprocal of the impedance and is given by 
\begin{equation}
    Y=1/Z
\end{equation}
Often, the Cole-Cole model serves as a description of how tissue impedance behaves at different frequencies. It is given by:
\begin{equation}
Z_{CC} = R_0 \left(1 + \frac{1}{1 + (\mathbf{i} \omega \tau)^{\alpha}}\right)
\end{equation}
where $Z_{CC}$ is the impedance given by the Cole-Cole model\cite{said2009variation}, $R_0$ denotes the DC resistance, relaxation time is given by $\tau$, and $\alpha$ is the shape parameter.\\
These physical properties of tissue play a crucial role in understanding and interpreting bioimpedance measurements. By considering the resistivity, conductivity, permittivity, impedance

\subsection{Bioimpedance based equations for Total Body Water (TBW)}
In recent years, BIA has evolved significantly to aid in assessment of body composition parameters. This section provides with a Table (\ref{tab:tbw equations}) of validated equations that use impedance measurements to estimate TBW. However, it is important to emphasize that all these equations are empirically derived and, consequently, are population-dependent. This dependency necessitates careful consideration when applying these equations to subjects whose demographic characteristics differ from those in the original validation groups. Here, sex is coded as male = 1 and female = 0.

\section{Chips and Devices}
\label{sec:cd}
A bioimpedance system typically consists of the following components: a voltage signal excitation, a voltage-to-current converter to connect to the bioimpedance sensor's current injecting electrodes, a voltage measurement front-end connected to the sensor's pertinent electrodes, and the back-end analog or digital processing of the recorded signal. Individual building blocks of commercially available chips can be used to manufacture instruments, and full-custom CMOS chip designs can be used to generate optimized and application-specific solutions. However, complete impedimetric commercial chip solutions also exist\cite{kassanos2021bioimpedance}. Based on our review of commercially available bioimpedance measurement solutions, we have identified the following chips, AD5933, ADuCM35x, AD5940/AD5941, ADAS1000, AFE4300, AFE4500, AFE4960, ams AS7058, MAX30001G, MAX30009 and MAX86178.
Out of these, eight key integrated circuit platforms that offer varying capabilities for impedance measurement applications. Table \ref{tab:bioimpedance_chips} provides a comprehensive comparison of these chips. This comparison can serve as a guide for researchers and developers selecting the appropriate chip according to their specific requirements for biomedical monitoring, wearable health devices, or clinical diagnostic systems.

\begin{table}
\centering
\caption{Comparison of Bioimpedance Measurement Chips}
\label{tab:bioimpedance_chips}
\resizebox{\columnwidth}{!}{%
\begin{tabular}{|l|c|c|c|c|}
\hline
\textbf{Chip Name} & \textbf{AD5933/34} & \textbf{ADuCM35x} & \textbf{AD5940/41} & \textbf{AFE4300} \\
\hline
\textbf{Frequency Range} & 1KHz to 100 KHz &  80 Hz to 75 KHz & 0 Hz to 200 kHz & 1 kHz to 255 kHz \\
\hline
\textbf{Resolution (ADC)} & 12-bit ADC, 1 MSPS & 12-bit ADC, 160 kSPS & 16-bit ADC, 800 kSPS & 16-bit ADC, 860 SPS \\
\hline
\textbf{Electrode Configuration} & \begin{tabular}[c]{@{}c@{}}2 electrodes\\ (4 with additional circuitry)\end{tabular} & Not specified & 4 electrode & \begin{tabular}[c]{@{}c@{}}Upto 3 sets of\\ 4 electrode configuration\end{tabular} \\
\hline
\textbf{Interface Protocols} & I2C & SPI, I2C, and UART & SPI & SPI \\
\hline
\textbf{Modalities} & \begin{tabular}[c]{@{}c@{}}Bioimpedance\\ Electrochemical Analysis\end{tabular} & Bioimpedance & \begin{tabular}[c]{@{}c@{}}Bioimpedance\\ Electrochemical Analysis\end{tabular} & \begin{tabular}[c]{@{}c@{}}Bioimpedance\\ Weight measurement\end{tabular} \\
\hline
\textbf{Relevant Publications} & \begin{tabular}[c]{@{}c@{}}Thoracic related:\cite{aqueveque2020simple, iqbal2024wearable, qiu2022wearable, pino2019wireless} \\ Other:\cite{kluza2025assessment, krishnan2024non, al2024diagnosis, zhang2024multi, muvsikic2024usage, khan2023internet, deshpande2023freshness} \end{tabular} & \begin{tabular}[c]{@{}c@{}} Other: \cite{alejandrino2023bioimpedance,guo2025construction,alejandrino2023detection}\end{tabular} & \begin{tabular}[c]{@{}c@{}}Thoracic related: \cite{sanchez2022wearable, jung2021impedance, berkebile2021towards, maravilla2023thoracic}\\ Other:\cite{tran2024design,search2024improving, liu2024ieat, liu2024imove, liu2024towards, hughes2023detection, asiain2022mswh} \end{tabular} & \begin{tabular}[c]{@{}c@{}}Thoracic related: \cite{john2020confirming}\\ Other:\cite{sulla2019non,fernandez2023low, bandur2023designing} \end{tabular} \\
\hline
\hline
\textbf{Chip Name} & \textbf{AFE4960} & \textbf{MAX30001G} & \textbf{MAX30009} & \textbf{MAX86178} \\
\hline
\textbf{Frequency Range} & 30 KHz to 100 KHz & 125 Hz to 131 kHz & 16 Hz to 500 KHz & 16 Hz to 500 KHz \\
\hline
\textbf{Resolution (ADC)} & 24-bit ADC & 20-bit ADC & 20-bit ADC, 4 kSPS & 20-bit ADC \\
\hline
\textbf{Electrode Configuration} & Not specified & 2 or 4 electrode & 2 or 4 electrode & 2 or 4 electrode \\
\hline
\textbf{Interface Protocols} & SPI and I2C & SPI & SPI and I2C & Not specified \\
\hline
\textbf{Modalities} & \begin{tabular}[c]{@{}c@{}}Bioimpedance\\ ECG\\ PPG \end{tabular} &  \begin{tabular}[c]{@{}c@{}}Bioimpedance\\ Galvanic Skin response\end{tabular} &  \begin{tabular}[c]{@{}c@{}}Bioimpedance\\ Galvanic Skin response\\ Impedance Cardiography and Plethysmography\end{tabular} & \begin{tabular}[c]{@{}c@{}}Bioimpedance\\ ECG\\ PPG \end{tabular}  \\
\hline
\textbf{Relevant Publications} & \begin{tabular}[c]{@{}c@{}}Not Available\end{tabular} &  \begin{tabular}[c]{@{}c@{}}Thoracic related:\cite{jain2025enhanced,yun2024towards} \\ Other:\cite{imbriglia2024assessment, plonis2024non, magno2020infiniwolf} \end{tabular} &  \begin{tabular}[c]{@{}c@{}}Thoracic related:\cite{mathews2023enabling} \\ Other:\cite{jovanov2024aerospace, hafid2023impact, crandall2022characterization, deng2024multichannel} \end{tabular} & \begin{tabular}[c]{@{}c@{}}Other: \cite{najafi2024versasens} \end{tabular}  \\
\hline
\end{tabular}%
}
\end{table}

\subsection{AD5933}
It comes with a high speed 12-bit analog-to-digital converter (ADC) and a frequency generator system for high accuracy impedance conversion. The system incorporates a Digital Signal Processing (DSP) engine that takes the sampled impedance response and calculates its discrete Fourier transform (DFT). The result of this DFT computation is real (R) and imaginary (I) values at various output frequencies. It can measure Impedance from 1 k$\Omega$ upto 10 M$\Omega$ and is also capable of measuring in the range of 100 $\Omega$ to 1 k$\Omega$ with help of an additional circuitry. Calibration of the system allows for the straightforward determination of both the relative phase and the impedance magnitude at each frequency measured during the sweep. This is done using the contents of the real and imaginary registers, read from the serial $I_2 C$ interface. However, this chip is not suitable for bioimpedance measurements at low frequencies. This is because the chip has a limited frequency range and resolution, and its accuracy decreases at low frequencies. Also, it is designed for two electrode measurements. It can be used with four electrodes but will require additional circuitry. Although it is configured for dual electrode configuration, Margo et al. designed a custom circuit to use it along with tetrapolar electrode settings to control the contact impedance between the electrodes and the material. They used the designed system for studying the conductivity of various physiological samples \cite{margo2013four}. Seoane et al. also designed an Analog Front End (AFE) and used it along the AD5933 circuit to fully adapt it to a four-electrode strategy, making it suitable for biomedical applications \cite{seoane2008analog}. It consisted of two voltage-to-current converters (V2CCs) and a current-to-voltage converter (CVC). The V2CCs convert the voltage signal from the AD5933 chip to a current signal, which is then applied to the Tissue of interest. The CVC converts the current signal from the biological tissue to a voltage signal, which is then fed back to the AD5933 chip for measurement. The maximum load dynamic range is determined by the overall gain introduced by the combined effect of both V2CCs and the resistor R$_{ref}$ at the input R$_{feedback}$. Overall, the AFE enables the use of AD5933 in biomedical applications by adapting it to a four-electrode strategy and providing a simple and effective way to convert voltage signals to current signals and vice versa.

\subsection{AD5940}
It is the newest iteration in the AD59xx series. It is specially designed for bio-impedance measurements.  It has two excitation loops along with a common measurement channel. The first excitation loop can generate signals from 0 Hz up to 200 Hz. Along with this, it also has a second excitation loop which can generate signals up to 200 kHz. It comes with a 16-bit (800kSPS) multichannel ADC along with input buffers. Featuring a built-in antialiasing filter and a programmable gain amplifier (PGA), the system enables the user to select the specific measurement input channel using an integrated input multiplexer (mux) positioned prior to the ADC. The sensor is connected to the internal analog excitation and measurement blocks using a programmable switch. Beyond impedance measurements, this chip also functions as a connection point for external trans-impedance amplifier resistors and calibration resistors. Furthermore, its programmable matrix switch allows multiple electronic measurement devices to share the same electrodes. To ensure stable performance with minimal drift for its 1.82 V and 2.5 V components, the chip incorporates an internal reference source. The serial peripheral interface (SPI) allows for direct register writes, which can be used to control the system. Alternatively, it can also be done through a pre-configurable sequencer for independent operation of the analog front-end (AFE) chip. Measurement instructions and outcomes are stored in separate command and data FIFOs. Various FIFO-related interrupts signal when the FIFO is full. The AFE sequencer also manages multiple general-purpose input/output pins (GPIOs) \cite{devices2020high}. In a study by Dutt et. al, the AD5940 sensor was utilized as a part of the bioimpedance measurement system for fluid monitoring in Dengue disease\cite{guru2020bio}. The EVAL-ADuCM3029 microcontroller development board, through the SPI protocol, was responsible for configuring the AD5940 sensor digitally to enable 4-wire whole-body bioimpedance measurements. The authors used it to generate a maximum output voltage of 1.2V (peak to peak) and limited the alternating current (AC) amplitude to a maximum of 400$\mu$A (rms) using a current-limited resistor. The AD5940 sensor performs measurements from 1 kHz to 100 kHz with 1 kHz increments at a rate of 20Hz. It also utilizes a DFT accelerator to determine the real and imaginary components of the measured bioimpedance. The stability and accuracy of the AD5940 single-chip converter were validated through capacitor test measurements in a study by Zhivkov et al. In the capacitor test measurements, the AD5940 was used to measure the impedance of capacitors at a single frequency of 1 Hz\cite{zhivkov2020detection}. The stability of the converter was evaluated by monitoring the impedance values over time. A stable converter would exhibit consistent impedance measurements without significant fluctuations or drift. The accuracy of the AD5940 was assessed by comparing the measured impedance values with the known values of the capacitors used in the test. AD5940 maintained consistent impedance measurements over time, indicating good stability and accurate operation. These stability and accuracy validation tests provide confidence in the performance of the AD5940 single-chip converter for impedance microbiology studies. In a study by Xia et al., it was used for ankle impedance analysis, interfaced with a microcontroller enabling impedance meaasurements ranging from 0 to 2000 Hz \cite{huang2025study}. The system utilized a programmable AC voltage generator, operating at a frequency of 50 kHz, with the AC amplitude restricted to a maximum of 500 $\mu$A (rms) via a current-limiting resistor. This configuration provided measurement accuracy to two decimal places, making it particularly suitable for clinical applications such as monitoring lower extremity edema in patients diagnosed with heart failure (HF). In another study presented by Asiain et al., it was used to design a wearable device for measuring electrodermal activity (EDA) in a constant alternating current setup \cite{asiain2025wearable}. The microcontroller, specifically the MAX32630 FTHR kit, interacted with the AD5941 via the serial peripheral interface (SPI) to facilitate impedance measurements. This allowed for the generation of high resolution voltage signals and maintained a maximum current through the skin at 10$\mu$A. The AD5941 performed measurements over a frequency range of 200 Hz to 200 kHz, utilizing a dual measurement loop for enhanced precision in EDA signal capture. Furthermore, it incorporated a discrete Fourier transform (DFT) hardware block to process the signals, effectively enabling the extraction of complex skin impedance values, thereby improving the overall reliability of the EDA data collected by the wearable device.

\subsection{AFE4300}\label{formats}
The AFE4300 is a low-cost AFE featuring two distinct signal chains: one for body composition analysis and the other for weight measurement. A 6-bit digital-to-analogue converter (DAC) is used for the weight measurement, and an external resistor controls the gain of the instrumentation amplifier (INA). For ratiometric measurements, a circuit is also included that drives an external bridge/load cell at a fixed 1.7 V. A 6-bit, 1-MSPS DAC and an internal pattern generator are used to create the sinusoidal current for bioimpedance analysis. This sinusoidal current can be applied to the body by a voltage-to-current converter to determine the body composition eventually. A differential amplifier measures and rectifies the voltage across the terminals due to the impedance. The amplitude is extracted and measured by the 16-bit ADC \cite{instruments2012datasheet}. In a study by Dahlmanns et al. AFE4300 serves as the analog front-end for the system for tetrapolar bioimpedance spectroscopy (BIS) in bandages to enable long-term monitoring of tissue integrity, specifically for chronic wounds such as ulcus cruris. It generates excitation frequencies ranging from 1 kHz to 255 kHz. It injects a current of 375 $\mu$A into the body, ensuring that the current remains below 500 $\mu$A, even with a +20\% error, to prevent any health impairment \cite{dahlmanns2020hardware}. Since AFE4300 is designed to be able to handle upto 3 tetrapolar-electrode configurations for impedance measurements. Sanchez et al. evaluated its potential to be used for various biomedical applications. Along with its interesting features to be used as a body composition impedance meter, the authors used it to monitor ventilatory bioimpedance varying over time \cite{sanchez2013minimal}.

\subsection{MAX3000x}
The MAX3000x family is a bioimpedance (BioZ) AFE, designed for wearable applications in consideration. With their very low power consumption and extended battery life, they offer outstanding performance for applications in clinical and fitness domain. MAX30001 is a single biopotential channel that can detect pacemaker edges, measure heart rate, and provide electrocardiogram (ECG) waveforms. It also has a single bioimpedance channel that can measure breath rates based on bioimpedance changes in the thoracic region. For built-in self-test, EMI filtering, detection of DC lead-off, internal lead biasing, ultra-low power, and leads-on detection when it's on standby mode, the biopotential and bioimpedance channels have a wide range of calibration voltages. No significant transients are injected into the electrodes thanks to soft power-up sequencing. High input impedance, low noise, configurable gain along with a variety of (low/high)-pass filters and a high-resolution ADC are further features shared by both channels. The biopotential channel is DC coupled, has a fast recovery mode, can manage significant electrode voltage offsets, and can swiftly recover from overdrive situations like defibrillation and electro-surgery. The bioimpedance channel provides the flexibility for 2 or 4 electrode readings, works with common electrodes, and has an integrated programmable current drive. AC lead-off detection is also present in the bioimpedance channel. The study by Critcher et al. states that the MAX3000x is a promising option for wearable applications due to its integration, small size, and versatility\cite{critcher2021localized}. The authors mention that the resistance measurements have a relative error of less than 10\% and can measure relative alterations in the 250 m$\omega$ range. The recommended gain setting for resistance measurements is an 80 v/v PGA (Programmable Gain Amplifier) gain to maximize accuracy. The study mentions the presence of power modes, filtering, and the potential trade-offs associated with different operating modes. Overall, it provides a technical overview of the MAX3000x integrated circuit and its potential applications in wearable systems. In another study by the same author, it's shown that the prototype system based on MAX30001 can measure more than 4 test impedances simultaneously\cite{critcher2020multi}. It integrates  multiplexers controlled by a microcontroller that enables measurements of 4 different test impedances by routing the excitation/measurement signals to multiple sites without the need of manual switching. This allows for efficient and simultaneous measurement of multiple test impedances using the MAX30001 integrated circuit. The primary objective of this study is to assess the level of accuracy achieved by the MAX30001 in impedance measurements. The prototype system allows for measurements of resistors and 2R-1C configurations with a frequency range from 1 kHz to 128 kHz. It also offers improved resistance-only measurements and features a two-point calibration to reduce measurement errors. 

\subsection{ADuCM35x}
The ADuCm355 SoC (System-on-Chip) device from Analog Devices is a versatile microcontroller that integrates various functionalities for sensor-based applications. It combines a high-performance ARM Cortex-M3 core with a precision analog front end, making it suitable for sensor measurement and control tasks. It features a 12-bit ADC along with a programmable gain amplifier (PGA), allowing it for flexible and accurate analog signal acquisition. It also includes a 12-bit digital-to-analog converter (DAC) for generating analog waveforms. In a study by Chen et al., the ADuCm355 is part of the sensing board used in conjunction with the textile based device for body sweat monitoring\cite{chen2021bio}. It is utilized to generate a sinusoidal wave and measure the impedance of the sensor connected to it. By analyzing the amplitude and phase shift of the wave, the device can determine the impedance. Additionally, the sensing board in this study incorporates other components, such as the FT232 USB interface IC from FTDI and the RN4871 BLE module from Microchip. These components facilitate communication and data transfer between the ADuCm355 device and external devices, enabling wireless connectivity for real-time monitoring. In another study, the ADuCM350 microcontroller is used for the development and characterization of a portable device called the Drug Under Skin Meter (DUSM) for measuring the volumne of drugs delivered through the transdermal delivery route\cite{arpaia2018bioimpedance}. The device utilizes bioimpedance measurements to assess the volume of the drug. It has an integrated system for impedance measurement, including magnitude and phase, and employs a parameterizable wave generator based on a 12-bit DAC. The on-chip 16-bit ADC is used to convert the current and voltage measurements. The ADuCM350 also performs a discrete Fourier transform to calculate the real and imaginary parts of the measured impedance.

\section{Application: Thoracic Fluid Detection}
\label{sec:app}
\subsection{Causes of thoracic fluid accumulation}
Thoracic fluid accumulation can result from various factors, including infections, lung disorders, kidney diseases, autoimmune conditions, and physical injuries. Nevertheless, the primary contributor to thoracic fluid buildup, known as pleural effusion, is congestive heart failure (CHF) \cite{karkhanis2012pleural}. CHF arises when the heart faces difficulty to pump blood efficiently, which leads to fluid retention in multiple parts of the body, including the pleural space and lungs. Elevated pressure within the blood vessels can cause fluid to seep into the thoracic cavity, ultimately causing pleural effusion  \cite{ponikowski2014heart}. 

The Starling Equation is commonly used in the evaluation of cardiogenic pulmonary edema to elucidate how oncotic and hydrostatic pressures contribute to net fluid buildup in the lungs \cite{levick2004revision}. It is expressed as:\footnote{Here, (trans endothelial solvent) filtration vol. per sec is denoted by $J_\nu$, hydraulic conductivity of the membrane by $L_p$, filtration surface area by S, capillary and interstitial hydrostatic pressure by $P_c$ and $P_i$, respectively. $\sigma$ is the Staverman's reflection coefficient and $\pi_p$ and $\pi_i$ are the plasma protein and interstitial oncotic pressure, respectively.}
\begin{equation}
    J_\nu=L_p S([P_c - P_i]-\sigma [\pi_p - \pi_i])
\end{equation}\\
According to this analysis, failure in the left ventricular results in fluid overload within the pulmonary circulation. Hydrostatic pressure increases, and the interstitial space in the lung, along with the alveoli, gets filled with fluid from capillaries. However, the Starling Equation does not fully explain this condition, otherwise, it would have rapidly resolved with standard medical interventions aimed at reducing the left ventricle's afterload and preload \cite{KRADIN2017297}. It was found that some hospitalized patients with this condition continue to experience respiratory symptoms and fluid accumulation even after receiving medical treatment \cite{pavlakis1997fatal}. Congestive heart failure is a widespread condition, particularly among the elderly, and remains a leading cause of hospital admissions. Although infections and lung disorders are common causes of thoracic fluid accumulation, they may not be as prevalent as CHF. Symptoms associated with the fluid overload that leads to heart failure-related hospitalization often appear very late, usually only after the interstitial fluid has increased more than six times the normal levels \cite{yu2005intrathoracic}\cite{shochat2006internal}. Hospitalizations related to heart failure frequently occur due to fluid overload \cite {yu2005intrathoracic}. Treating acute heart failure during its preclinical phase can help prevent its progression or reduce its clinical impact. To ensure timely patient care and reduce hospitalizations, healthcare professionals must be adept at identifying this preclinical phase, with a key component being the assessment of increased thoracic fluid content \cite{shochat2006internal}\cite{abraham2007intrathoracic}. Thoracic fluid content is an umbrella term used to infer fluid present in the intravascular, interstitial, and intra-alveolar regions of the lung. Monitoring the amount of fluid in the chest during potential heart failure worsening can help differentiate between cardiac-related shortness of breath and other causes. It's crucial to note that thoracic fluid content is only a portion of a patient's total fluid balance \cite{folan2008measurement}. Clinical guidelines from the American College of Cardiology and the American Heart Association in 2005 recommend observing alterations in body weight, orthostatic blood pressure, jugular venous distension, hepatojugular reflex, the degree of organ congestion, edema in various locations (legs, abdomen, presacral area, scrotum), and abdominal ascites. However, it is noticed that weight change can be affected by muscle atrophy or hypertrophy and other physical factors; hence, it may not exclusively indicate an increase in fluid levels. Distinguishing thoracic fluid changes from overall body weight changes can be challenging \cite{davies2014role}. All these evaluation criteria are subject to variability based on the healthcare professional conducting the assessment.
\subsection{Conventional methods}
There are multiple ways to check for fluid accumulation in the lungs, such as pulmonary artery wedge pressure\cite{connolly1954relationship,fink1986exercise,oldenburg2009pulmonary}, chest x-ray\cite{matsumoto2020diagnosing,pan2021prognostic,costanzo1988role}, Echocardiography\cite{kirkpatrick2007echocardiography,marwick2015role,wheeldon1993echocardiography}, B-type Natriuretic Peptide (BNP) blood test\cite{sartini2017method}, Computed Tomography (CT), MRI, etc. Here, we will discuss a few of them, which are frequently used as standard. 
\subsubsection{X-ray Imaging:} The chest X-ray remains a crucial diagnostic tool for assessing acute pulmonary conditions, offering valuable insights into differentiating between cardiogenic pulmonary edema and non-cardiogenic pulmonary edema \cite{dobbe2019cardiogenic}\cite{gluecker1999clinical}. Milne et al. identified key radiological indicators to aid in the detection of cardiogenic pulmonary edema. These indicators encompass an altered distribution of how the blood flows within the lungs, a uniform and basal spread of fluid from the chest wall to the heart, either normal or reduced lung volumes, as well as the presence of  septal lines, peribronchial cuffing and effusions \cite{milne1985radiologic}. Certain patterns observed in chest X-rays can indicate the reasons behind edema. For instance, cardiomegaly and pleural effusions are frequent findings on X-rays when the cause of pulmonary edema is related to the heart, thus aiding in determining the primary issue. However, it's worth noting that the features visible in chest X-rays exhibit moderate specificity (ranging from 75\% to 83\%) and limited sensitivity (ranging from 50\% to 68\%) when diagnosing pulmonary edema \cite{mant2009systematic}. In another study, Hammon et al. proposed a standardized scoring procedure to enhance the accuracy of diagnosing pulmonary edema identification in chest radiographs. Their study involved an equal number of patients with and without pulmonary edema. Radiologists assessed the X-rays using their conventional methods or custom software designed to outline pulmonary edema characteristics and assigned scores for the abnormalities detected in the images. This approach required around 20 seconds per image and significantly improved sensitivity and specificity for detecting pulmonary edema from 57\% to 77\% and 90\% to 100\%, respectively \cite{hammon2014improving}.
\subsubsection{Chest CT:} Mooroka et al in 1982 performed a study in which a region of interest was selected to avoid larger blood vessels in the one-third of both anterior and posterior lung regions \cite{morooka1982estimation}. It was found that the mean CT value (Hounsfield Units) of the anterior and posterior lung region was -432.1 HU, 8.1 HU in 26 normal subjects, -370.6 HU, 14.0 HU in the group diagnosed with interstitial edema,  -392.3 HU, 15.1 HU in the pulmonary congestion group and -349.2 HU, 27.4 HU in the group diagnosed with alveolar edema. In this study, the authors found the mean CT value to be higher for the posterior lung field compared to the anterior lung field in the supine position, however, these findings get reversed when in the prone position. Kato et al also performed a similar study in which they analyzed the CT images using an analyzer (Toshiba GMS-55U) to get CT numbers along with histograms of those CT numbers \cite{kato1996early}. It was found that in normal people the values were in the range of –950 HU to –650 HU. It was a comparative study in which weighted mean CT number of patients diagnosed with severe heart failure were compared with that of a group of healthy individuals. The histograms made an intersection at around –750 HU. Since CT numbers from –750 to –650 in the histograms represent the patients with heart failure conditions, a term \%pixel was defined as the ratio of pixel number of CT number from -650 to -750 HU in the pulmonary ROI and the pixel number of CT number from -300 to -950 HU in the same pulmonary region, this ratio was then multiplied by 100 to get the \%pixel. Simon et al discussed and showed how CT numbers can be used to determine the volume of the desired area and came up with the following equation,  
\begin{equation}
    Density(ROI) = \frac{(D_{air}*V_{air} +D_{tissue}*V_{tissue})}{(V_{air} + V_{tissue})}
\end{equation} 
\begin{equation}
    V(ROI) = V_{air} + V_{tissue}
\end{equation}
where Density(ROI) is the density of the volume of the lung measured in the CT scan (the `region of interest' (ROI)), $V_{air}$ and $V_{tissue}$ the ejective volumes of the air and tissue compartments of the ROI, and $D_{air}$ and $D_{tissue}$ the CT densities of air and tissue. Density(ROI) and V(ROI) are measured from the CT image, and $V_{air}$ and $V_{tissue}$ are calculated readily from the above two relationships \cite{simon2000non}. \\
In another study by Snyder et al, Pulmonary Analysis Software Suite was used to process CT images to analyze lung density and the volume of the tissues \cite{snyder2007genetic}. The CT images were fed to the software and it segmented the lung tissue from the surrounding structures. In this context, lung density was modeled as a linear mix of the Hounsfield Units (HU) for air (-1000) and lung tissue (0). The lung density and tissue volume data were obtained using linear interpolation of the plotted data from baseline images. In this study, the following equation was used to estimate the changes in the lung water from baseline to post-saline: 
\begin{equation}
    Lung\:water = (TV_{Pst}-Vc_{Pst})-(TV_{IP}-Vc_{IP})
\end{equation}
For this analysis, $TV_{Pst}$ is the post-intervention CT scan tissue volume (in ml), and $TV_{IP}$ is the interpolated tissue volume (in ml) at the post-intervention lung volume derived from the baseline CT scans and $Vc_{Pst}$ and $Vc_{IP}$ are the Vc(pulmonary capillary blood volume) (in ml) at post intervention and at baseline time, respectively. Here, the tissue volume and Vc were obtained by using the tests designed for measuring lung diffusing capacity.\\
However, in a recent study, Chest CT scans were used to validate a medical device (ReDS system) result \cite{amir2016validation}. In this study, the authors defined Mean Lung Density (MLD) to be the average attenuation value (HU) of all lung voxels. It was used to determine the total fluid content in the lungs. This value was converted to a fluid\% by using the following equation, 
\begin{equation}
    F[\%] = \frac{(MLD + 1000)}{10}
\end{equation}
Due to variations in lung air volume, while holding breaths, the fluid\% values were adjusted for the mean lung volume. 
                               
\subsection{Bioimpedance based studies}
Using impedance monitoring of the thoracic region offers several benefits compared to traditional diagnostic techniques. Firstly, it provides early detection of fluid accumulation in the lungs, allowing for timely intervention and prevention of heart failure decompensation. This can help reduce the incidence of hospitalizations and improve patient outcomes Secondly, intrathoracic impedance monitoring can be performed continuously and remotely, providing real-time updates to healthcare providers. This allows for proactive management of heart failure patients, as changes in impedance can be detected even before symptoms become apparent. Additionally, by helping provide adequate in-patient diuresis, this technology can help guide therapy during hospitalization, potentially reducing the chance of re-admissions \cite{abraham2007intrathoracic}. In recent times, cardiac defibrillation and re-synchronization devices have integrated intrathoracic impedance monitoring systems, exemplified by the OptiVolTM fluid index from Medtronic Inc. These systems detect the drop in the intrathoracic impedance, signaling potential fluid accumulation in the thoracic region and alerting patients to potential cardiac decompensation \cite{vollmann2007clinical,catanzariti2009monitoring}. Current research data suggests that monitoring of the intrathoracic impedance holds promise for early detection of chronic heart failure (HF) \cite{adamson2004continuous,capucci2019preliminary}. Alternatively, non-invasive measurements of trans-thoracic electrical impedance have been proposed as a reliable and patient compliant means to assess fluid shifts within the thoracic cavity \cite{pennati2023electrical,groenendaal2021wearable}. In a study conducted by Chu-Pak et al., patients diagnosed with NYHA class III and IV heart failure had bioimpedance sensors attached to measure intrathoracic impedance \cite{yu2005intrathoracic}. The study investigated three stimulation/measurement pathways for assessing intrathoracic impedance. In the methodology they followed, constant current injection to the biological tissue happened at a frequency of 16 Hz, this was independent of the cardiac rhythm. Simultaneously, the voltage drop was recorded from another pair of electrodes to calculate the intrathoracic impedance. The data was collected for 2 minutes and then averaged out. The 2-minute time period was chosen to reduce the effects due to respiration and cardiac rhythm. For the chronic patients, data was collected after every 45-minute window for the next 6 hours. The study established a strong connection between two factors, intrathoracic impedance and pulmonary capillary wedge pressure. Continuous monitoring of intrathoracic impedance for patients undergoing acute heart failure treatment offered quantitative insights into the severity of fluid volume overload. Venegas et al. aimed to explore the potential role of Bioelectrical Impedance Analysis (BIA) as an dditional tool to diagnose Acute Decompensated Heart Failure (ADHF) \cite{venegas2023role}. Their objective was to evaluate the benefits of BIA in managing fluid overload in HF patients. The study randomized 48 patients, with 24 patients receiving standard clinical treatment and the other 24 receiving BIA-guided treatment. The findings indicated that BIA is a non-invasive, cost-effective, and reproducible method for estimating body mass and water composition, offering valuable insights into the detection of peripheral congestion. The study acknowledged limitations in sample size and the pilot nature of the research, emphasizing the need for larger multicenter trials with extended follow-up periods to validate these findings. The study suggested that BIA can assist in determining euvolemia in heart failure patients by assessing fluid overload, thereby providing a more accurate diagnosis and supporting informed decisions regarding diuretic therapy. Furthermore, the study underscored the potential advantages of BIA in enhancing the management of patients presenting with acute respiratory distress in the Emergency Department, enabling faster and more accurate diagnoses, and aiding in the decisions related to diuretic therapy. In conclusion, the study positioned BIA as a promising tool in managing fluid overload in heart failure patients, offering valuable insights into the detection of peripheral congestion. Remote dielectric sensing (ReDS) is another non-invasive technology that uses a high frequency electromagnetic signal transmitter and receiver module, applied to the front and back of the chest to calculate absolute lung fluid volume eventually. This study by Sattar et. al found that monitoring fluid changes using this device significantly decreases the chances of readmission cases of HF within three months of their hospitalization \cite{sattar2021efficacy}. Similarly, Bensimhon et al. also aimed to evaluate the use of this device to guide discharge decisions in patients hospitalized with acute HF \cite{bensimhon2021use}. The study enrolled 130 patients, randomized to treatment or control arms, and measured ReDS readings on the day of proposed discharge. The primary endpoint was the percentage of patients with evidence of significant persistent fluid overload despite planned discharge. The study found that 32\% of patients had significant persistent fluid overload, and an additional 12\% had mild lung congestion at discharge. In the treatment group, patients with initial ReDS readings $>$39\% were considered high-risk and referred to the outpatient HF clinic for follow-up. When compared to the control group, where initial readings were also $>$39\%, the ReDS-guided approach led to a significant decrease in heart failure readmissions, dropping from 11.8\% to 0\% at 30 days and from 23.5\% to 9.1\% at 90 days. The study concluded that the use of ReDS technology could potentially improve patient outcomes and reduce healthcare costs by optimizing HF management and reducing the need for hospital readmissions. However, the study had several limitations, including single-center enrollment, a small sample size, and a lack of prospective validation of the 39$\%$ cutoff point.

\section{AI for Bioimpedance analysis}
\label{sec:ai}
In this section, we delve into the application of Artificial Intelligence (AI) and Machine Learning (ML) techniques within the domain of bioimpedance. These cutting-edge technologies have revolutionized our approach to cardiovascular diagnostics and fluid dynamics analysis. We explore specific studies that exemplify how AI and ML enhance the precision and insights gained from bioimpedance measurements, bridging computational intelligence with biomedical research to advance early disease detection and patient care.\\
In a study by Dovancescu et al., around 1300 ECG recordings from over 35 volunteers were collected for electrocardiography and transthoracic bioimpedance data \cite{dovancescu2015detecting}. After the participants were discharged from the hospital, their health condition was monitored using a bioimpedance enabled vest. 5 minute long, multiple time series ECG signals were recorded using the vest. The transthoracic bioimpedance data along with ECG collected in this research study were aimed to predict acutely decompensated heart failure (ADHF. The researchers used a multilayer perceptron (MLP) regression model on the ECG data to predict transthoracic bioimpedance. They then used the predicted bioimpedance values as an additional feature in their deep-learning models to predict ADHF. The participants wore the vest during routine/daily procedures, and the raw time series data was processed to remove poor-quality ECG recordings and normalization was done to use the data for cross-modal feature learning and deep learning models. The results showed that the addition of bioimpedance data improved the performance of the models, indicating that bioimpedance data can be a useful feature in predicting ADHF. The researchers also used bioimpedance data to perform a regression analysis to estimate the fluid accumulated in the intrathoracic region. They used type 1 data (ECGs data along with bioimpedance data) for the training and validation set and the type 2 data (ECGs without bioimpedance values) as the test set. The results showed that the MLP regressor achieved a mean absolute error (MAE) of 0.23 L, indicating that bioimpedance data can be used to estimate the intrathoracic fluid accumulation level accurately. Overall, the results associated with bioimpedance data in this research demonstrate its potential as a useful feature in predicting ADHF and estimating the intrathoracic fluid accumulation level \cite{pan2023deep}. In a different study conducted by Jonathan Moeyersons et al., the authors focused on the use of a machine-learning based approach to detect artifacts in thoracic impedance signals. The study evaluated and compared three distinct methods: a heuristic approach and two ML approaches \cite{moeyersons2021artefact}.The first ML-centric approach utilized a Support Vector Machine (SVM) classifier, where distinctive features representing various signal aspects were manually crafted. These features were subsequently input into a classification model, segregating them into different quality classes. The other approach involved using a Convolutional Neural Network (CNN) that directly learned characteristic features from the raw data and used them for classification. The study's results demonstrated a significant improvement in accuracy for both machine learning approaches, with SVM achieving an accuracy of 87.77\% ± 2.64\% and a CNN achieving 87.20\% ± 2.78\%. In contrast, the heuristic approach achieved an accuracy of 84.69\% ± 2.32\%. Furthermore, the feature-based and neural network models exhibited respective Area Under the Curve (AUC) values of 92.77\% ± 2.95\% and 92.51\% ± 1.74\%. These findings demonstrate the value of a data-driven strategy for the essential process of identifying artifacts within thoracic bio-impedance signals obtained during respiratory monitoring.\\
AI, when incorporated with impedance also benefited in identification of lung neoplasm. A recent study by Nescolarde et al. used electrical impedance spectroscopy to assess the electrical properties of lung tissues at multiple frequencies, ranging from 15 KHz to 307 KHz \cite{company2025machine}. Neoplastic and non-neoplastic tissues showed significant differences in the impedance values. ML algorithms such as Decision Tree, Discriminant Analysis, Support Vector Machines (SVM) along with K-Nearest Neighbors (KNN) and Ensemble Method were trained to differentiate between the different types of tissues. Most of the models achieved accuracy level higher than 95\%, highlighting ML's potential to enhance the diagnostic accuracy and reduce reliance on invasive methods.
\section{Challenges and limitations}
\label{sec:chal}While bioimpedance measurements offer promising applications in various fields, including body composition analysis and thoracic impedance monitoring for heart failure, several technical challenges need to be addressed for accurate and stable readings. In this subsection, we will discuss these challenges and propose potential solutions to overcome them. 
\subsection{Electrode-Skin Interface impedance}
One of the primary challenges in bioimpedance measurements is ensuring a reliable and stable electrode-skin interface. Measurement accuracy is highly dependent on how well the electrodes are in contact with the skin. Poor contact can introduce impedance artifacts and noise, leading to erroneous readings. For bio-impedance measurements, electrodes can be configured either in a bipolar or tetrapolar configuration. The bipolar configuration involves the utilization of 2 electrode (single pair) to apply an electrical current and simultaneously measure the resulting potential drop. This approach was originally introduced by Thomasset in 1963 when he conducted pioneering research into estimating total body water through electrical impedance measurement. Needle-type electrodes were employed for this purpose \cite{ma1962bioelectric}. This led to a straightforward circuit and overall system configuration due to its minimal electrode requirement. However, its precision is compromised by contact impedance issues. To mitigate measurement errors attributable to contact resistance, a technique involving the use of tetrapolar electrode configuration was developed by Hofer et al. and Nyboer \cite{hoffer1969correlation,nyboer1950electrical}. Their method incorporates use of 4 electrodes (two pairs), distinctly allocating the electrodes responsible for applying the current from those that measure the voltage. One pair is designated for delivering electrical current into the tissue of interest, while the other pair of electrodes is responsible for detecting the voltage drop across it. Jongae Park et. al. proposed another novel contact resistance compensation method to accurately measure bioimpedance values even with the small electrodes \cite{jung2021wrist,jung2016wrist}. 
\subsection{Motion Artifacts}
Another significant challenge in bioimpedance measurements is dealing with motion artifacts. Movement during the measurement can introduce noise and distort the impedance signal, making it difficult to obtain reliable data. To lessen the impact of movement-related noise, researchers are utilizing different signal processing methods, including adaptive filtering and artifact rejection algorithms. Additionally, advances in wearable technology and sensor integration enable real-time monitoring and motion compensation, allowing for continuous measurements even during physical activity. 
\subsection{Calibration and Validation}
Ensuring precise and dependable bioimpedance measurements is of utmost importance to obtain meaningful outcomes. The process involves accurate calibration to known reference values and validation against established gold standard methods. Standardizing calibration protocols and validation procedures is crucial for consistency and comparability across various bioimpedance measurement systems. Collaboration among researchers, clinicians, and regulatory bodies is essential in establishing guidelines and benchmarks to govern the calibration and validation processes.                       
\subsection{Interference and Noise}
Bioimpedance measurements are susceptible to interference from various sources\cite{heikenfeld2018wearable}, including electromagnetic noise and physiological signals originating from adjacent organs or tissues \cite{lin2022soft}. To minimize interference, careful system design and shielding techniques are employed to reduce external noise. Additionally, advanced signal processing methods, such as adaptive filtering, spectral analysis, and noise cancellation algorithms, are used to extract the relevant bioimpedance information while suppressing unwanted noise and interference \cite{martinek2021advanced}. 
\section{Conclusion}
\label{sec:con}
This systematic review article puts forward a comprehensive assessment of thoracic bioimpedance research within the realm of heart failure diagnosis. 
Significant advancements in sensor technology have yielded highly precise and compact sensors that redefine monitoring strategies. Thoracic bioimpedance emerges as a promising non-invasive alternative for detecting thoracic fluid accumulation associated with chronic heart failure cases, addressing the limitations posed by traditional methods like X-ray and CT scans. 

Notably, the integration of machine learning algorithms and artificial intelligence plays a central role in signal processing and data analysis within this field. This technological synergy holds immense potential for extracting crucial insights from bioimpedance data, promising enhanced diagnostic precision and more informed treatment decisions. As technology continues to evolve, the prospects for improved diagnostic accuracy and widespread clinical adoption of thoracic bioimpedance appear promising, heralding potential transformations in heart failure diagnosis and management. 

\bibliographystyle{IEEEtran}
\bibliography{reviewref}

\end{document}